 \documentclass[final,5p,times,twocolumn]{elsarticle}

\usepackage{amsmath,amssymb,amsfonts}
\usepackage{mathrsfs}
\usepackage{chemformula}
\usepackage{color}
\usepackage[normalem]{ulem}
\usepackage{stmaryrd}
\usepackage{diagbox}

 \biboptions{sort&compress}

\newcommand{\beq}{\begin{equation}}
\newcommand{\eeq}{\end{equation}}

\newcommand{\mb}[1]{\mathbf{#1}}

\newcommand{\gb}[1]{\boldsymbol{#1}}
\newcommand{\mc}[1]{\mathcal{#1}}

\journal{Combustion and Flame}

\begin{document}

\begin{frontmatter}

\title{Premixed flames in a stagnation point flow under Darcy's law}

\author{Prabakaran Rajamanickam and Joel Daou}
\address{Department of Mathematics, University of Manchester, Manchester M13 9PL, United Kingdom}

\begin{abstract}
Premixed flames in stagnation point flows are traditionally described using Navier--Stokes equations where inertia and density variations play an important part in determining the flame structure. However, in porous media or Hele-Shaw configurations, Darcy’s law replaces the momentum balance, shifting the governing physics to a balance between pressure and viscous forces. This study investigates non-adiabatic strained premixed flames under Darcy’s law, pertinent in particular to confined flames in Hele-Shaw burners, accounting for non-unity Lewis numbers and volumetric heat losses. The flame is established in a planar counterflow formed by impinging a cold unburnt gas and a hot burnt gas maintained at the adiabatic flame temperature. We show that the jump in the strain rate across the flame is associated with a jump in viscosity, rather than, as in the classical Navier–Stokes case, a jump in density. Furthermore, the ratio of viscosity to the density-permeability product $\mu/\rho \kappa$, i.e., kinematic viscous resistance, is identified as a key coordinate stretching factor in the mathematical description of the flame structure.  This ratio increases significantly across the flame. As a result: (1) the burnt gas acts as a strong viscous barrier, (2) for an increasing strain rate, flame migration towards the burnt gas is hindered, (3) for a decreasing strain rate, migration towards the unburnt gas is promoted,  and (4) streamline refraction is augmented. By analysing the burning rate across varying strain rates and heat-loss parameters, we identify distinct extinction and ignition regimes that fundamentally differ from classical combustion theory, thereby providing new insights into flame stabilisation in friction-dominated environments and under confinement.
\end{abstract}

\begin{keyword}
    Darcy's law \sep confined flames \sep Hele-Shaw burners \sep porous-media combustion \sep variable-property hydrodynamics 
\end{keyword}

\end{frontmatter}

\section{Introduction}

Stagnation point flows serve as a cornerstone in fluid mechanics and combustion science, providing a canonical framework for investigating the coupling between hydrodynamic strain, molecular transport, and chemical kinetics. Traditionally, these flows are analysed using the Navier--Stokes equations in the high-Reynolds-number limit. When an interface such as a flame is present in such flows, the strain-rate jump across the interface is dictated by the density jump. In contrast, in porous media or Hele-Shaw configurations, Darcy's law fundamentally alters the physics: the momentum balance reduces to a competition between viscous and pressure forces with no inertia. As a result, the strain-rate experiences a jump across the flame which is associated with a jump in viscosity rather than density, a point briefly highlighted in our recent work~\cite{rajamanickam2026flame}. In the present study, we go further to demonstrate that, more generally, spatial variations in the strain rate are associated with variations in viscosity, with no direct contribution from density changes.  A general hydrodynamic theory of premixed flames under Darcy's law has been formulated in our recent works~\cite{rajamanickam2026hydrodynamic,rajamanickam2026flame,rajamanickam2024hydrodynamic,daou2025hydrodynamic}.

This study aims to elucidate the structure of nonadiabatic premixed flames governed by Darcy's law, which are relevant to flames in narrow Hele-Shaw burners and porous media. As such, it complements the extensive theoretical investigations~\cite{libby1982structure,buckmaster1982premixed,libby1983strained,libby1983strained2,daou2011strained,vera2023large}, numerical studies~\cite{darabiha1986effect,darabiha1988extinction,davis2002determination} and experimental studies~\cite{puri1987extinction,chao1997structure,smooke1991comparison}, based on the Navier--Stokes equations. Notable in the literature is the work by Libby, Liñán \& Williams~\cite{libby1983strained} which provided a detailed flame characterisation depending on the Lewis number and hot gas temperature. More recently, Vera \& Liñán~\cite{vera2023large} provided a detailed description of nonadiabatic strained flames for arbitrary Lewis numbers, identifying several distinct flame regimes. Particularly, while abrupt transitions in the burning rate $\dot m$ leading to the familiar S-shaped ignition--extinction curve in the $\dot m$--$Da$ plot have long been known for Lewis numbers $Le \gg 1$, Vera \& Li\~n\'an~\cite{vera2023large} recently reported such transitions also for $Le \ll 1$, relevant to hydrogen combustion. 

In the present study, we show that the premixed flame is governed by canonical equations with coefficients influenced by Darcy flow. The flame is established in a planar counterflow of cold unburnt gas against a hot inert gas. In contrast to~\cite{vera2023large}, where nonadiabaticity arises from prescribed variations in the hot inert gas temperature (not necessarily adiabatic), our model includes volumetric heat losses. In the case of Hele-Shaw burners, these losses arise from heat conduction at nonadiabatic walls, which manifest as volumetric terms in the depth-averaged equations.   For simplicity, and to isolate the impact of these volumetric losses, the impinging hot gas is assumed to be at the adiabatic flame temperature. 

\section{Preliminary considerations: A steady non-reacting mixing layer under Darcy's law}

To highlight the fundamental distinction between inertia-dominated and friction-dominated flows, we consider a steady, non-reacting mixing layer in a planar stagnation-point configuration. The flow is established by the impingement of two opposing streams with distinct physical properties, in the $xy$-plane, approaching from $y\to \pm \infty$ and exiting toward $x \to \pm \infty$.  Physical properties including density $\rho(\theta)$, viscosity $\mu(\theta)$, diffusivity $D(\theta)$ and medium permeability $\kappa(\theta)$ are coupled to a normalised scalar field $\theta\in[0,1]$, that characterises the mixing of the two fluids. In this paper, $\theta$ represents the dimensionless temperature defined as $\theta = (T - T_1)/(T_2 - T_1)$, where $T_1$ and $T_2$ are the temperatures of the upper and lower streams, respectively. The fluid density satisfies the ideal gas equation of state $\rho T = \text{constant}$, while the transport properties  vary with temperature according to a power-law  relation of the form $\mu/\mu_1 = \rho D/(\rho D)_1 =  (T/T_1)^n$, with $n=0.7$ as adopted in Section 3. For Hele-Shaw cells with a gap width $h$, the permeability $\kappa=h^2/12$ is a constant. More generally, $\theta$ can represent any other mixing scalar field such as solute concentration or salinity. We denote the properties of the upper and lower streams with subscripts  $1$ and $2$,  respectively. The governing equations are given by
\begin{align} \label{GovEqs}
    \nabla\cdot(\rho \mb v)  = 0, \quad -\frac{\mu}{\kappa} \mb v   = \nabla p, \quad  \rho \mb v \cdot \nabla \theta   = \nabla\cdot(\rho D \nabla \theta).
\end{align}
The steady mixing layer can be described by a self-similar solution of the form
\begin{align} 
    &v_x = A(y) x, \quad v_y = v_y(y), \quad \theta= \theta(y), \label{selfsimDarcy1}\\
    &p= p_1-\frac{A_1\mu_1}{2\kappa_1}(x^2-y^2) + \Pi(y), \label{selfsimDarcy2}
\end{align}
where $A(y)$ is the strain rate. Since $\rho$ and other transport properties are assumed to be a function only of $\theta$, they are also independent of the transverse coordinate $x$. Note that the self-similar solution is characterised by a pressure field quite different from that of a conventional counterflow, the latter being briefly addressed in the appendix.

Substituting the ansatz~\eqref{selfsimDarcy1}-\eqref{selfsimDarcy2} into the governing equations~\eqref{GovEqs} yields
\begin{align}
    A(y) &= A_1 \frac{\kappa/\mu}{\kappa_1/\mu_1}, \label{strain1} \\  \rho v_y &= - \frac{A_1\mu_1}{\kappa_1} \int_0^y \frac{\rho\kappa}{\mu} dy, \label{strain2} \\ \rho v_y \frac{d\theta}{dy} &= \frac{d}{dy}\left(\rho D \frac{d\theta}{dy}\right),   \label{strain3}
\end{align}
where the last equation needs to be solved using $\theta(+\infty)=0$ and $\theta(-\infty)=1$. The substitution also provides the governing equation and boundary condition for the pressure field $\Pi(y)$, associated with the mixing process:
\begin{equation}
    \frac{d\Pi}{dy} =-\frac{\mu}{\kappa}\left(v_y + A y\right), \qquad \Pi(+\infty)=0.
\end{equation}
The formal implicit solution for $\theta(y)$ is given by
\begin{align}
    \theta(y)&=1- \frac{f(y)}{f(+\infty)}, \quad \text{where}\\ f(y) &= \int_{-\infty}^y \frac{dy'}{\rho'D'} \exp\left(\int_{-\infty}^{y'}\rho''v_y''\frac{dy''}{\rho''D''}\right).
\end{align}

An important point to note is that, unlike under Navier--Stokes hydrodynamics, an exact expression for the strain rate $A(y)$ can be obtained under Darcy’s law, given in~\eqref{strain1}. The expression implies that there is a jump in the strain-rate between the two fluids given by
\begin{equation}
     \frac{A_1}{A_2} = \frac{\kappa_1/\mu_1}{\kappa_2/\mu_2}.
\end{equation}
This is fundamentally different from the conventional relation, $A_1/A_2 = \sqrt{\rho_2/\rho_1}$, depending on the density ratio, applicable under Navier--Stokes equations~\cite{weiss2017aerodynamics,linan2017large}.

\begin{figure}[h!]
\centering
\includegraphics[scale=0.6]{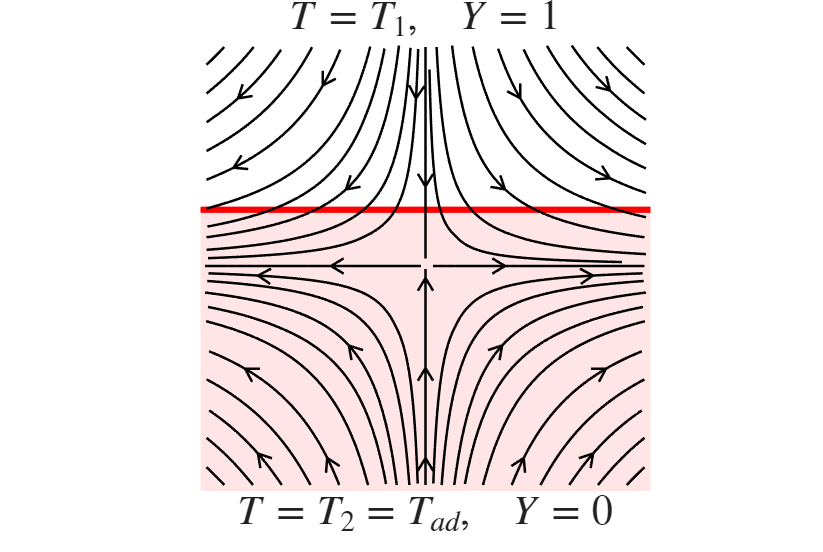} 
\caption{Schematic illustration of a strained premixed flame in a two-dimensional counterflow.}
\label{fig:sch}
\end{figure}

\section{Strained premixed flame in a planar counterflow: Formulation}

We now consider a strained premixed flame in a planar counterflow, formed by the impingement of a fuel-lean mixture against a hot  gas stream maintained at the adiabatic flame temperature,  as shown in Fig.\ref{fig:sch}. In this configuration, fluid 1 corresponds to an unburnt reacting mixture and fluid 2 to a burnt inert gas mixture. For simplicity,  the unburnt gas is assumed  to be deficient in a reactant which is consumed with a reaction rate  per unit volume  given by an  Arrhenius law $\rho B Y e^{-E/RT}$. Here, $B$ is the pre-exponential factor, $Y$ the reactant mass fraction,  $T$ the  gas temperature, $E$ the activation energy of the reaction, and $R$ the universal gas constant.   For this model, we may define the adiabatic flame temperature by $T_{ad}=T_1(1+q)$, where $q$ quantifies the amount of heat released by the reaction and $T_1$ the unburnt gas temperature. A non-dimensional measure of the activation energy is then given by  the Zeldovich number $\beta = E(T_{ad}-T_1)/RT_{ad}^2$. Conductive heat losses are modelled by the presence of a  heat-loss rate (per unit volume) term which takes the form $\rho_1 c_p K_h (T-T_1)$ as in Newton's law of cooling~\cite{daou2002influence,daou2011strained,martinez2019role}, with $c_p$ being the specific heat at constant pressure and $K_h$ a heat-transfer coefficient. The gas density  is assumed to depend on temperature according to the ideal gas law, while the transport coefficients are assumed to follow a power-law temperature dependence. Specifically, 
\[
\rho T = \rho_1 T_1 \,, \quad \frac{\mu}{\mu_1} = \frac{\rho D_T}{\rho_1 D_{T,1}} = \left(\frac{T}{T_1}\right)^n  
\]
with $n=0.7$. Since we assume that the incoming hot gas temperature is set to the adiabatic flame temperature, $T_2=T_{ad}$. The density ratio $r$ and viscosity ratio $m$ are defined by
\begin{equation}
    r = \frac{\rho_1}{\rho_2}=1+q, \qquad m = \frac{\mu_1}{\mu_2}=(1+q)^{-n}.
\end{equation}
Our investigation here focuses on Hele-Shaw burners for which the problem is two-dimensional and $\kappa_1=\kappa_2=h^2/12$, where $h$ is the channel width. 

For convenience, we introduce the following non-dimensional variables and parameters:
\begin{align}
    &\mb x^* = \frac{\mb x}{\delta_L}, \quad \rho^* = \frac{\rho}{\rho_1}, \quad \mb v^*= \frac{\mb v}{S_{L}}, \quad  \mu^* = \frac{\mu}{\mu_1}, \label{eqn1} \\ &p^* = \frac{h^2 p}{12\mu_1D_{T,1}}, \quad \theta = \frac{T-T_1}{T_2-T_1}, \quad Y^* = \frac{Y}{Y_1},\label{eqn2}  \\ &  \lambda = \frac{\rho D_T}{\rho_1 D_{T,1}},  \quad \mc A = \frac{A_1}{S_L/\delta_L}, \quad K =  \frac{\beta K_h\delta_L}{S_L}, \quad S= \frac{S_L}{S_L^0},\label{eqn3} 
\end{align}
where $\delta_L=D_{T,1}/S_L$ is the laminar flame thickness, $S_L$ is the burning speed of the unstrained planar flame computed numerically. This is expected to approach the theoretical value $S_L^0$ given by
$(S_L^0)^2=  2Le \beta^{-2} B  D_{T,2} (\rho_2/\rho_1)^2 \, e^{-E/RT_2}$, in the asymptotic limit $\beta\to\infty$.
 Dropping the asterisks for $\mb x^*$, $\rho^*$, $\mb v^*$, $p^*$, $Y^*$ and $\mu^*$, the two-dimensional governing equations can be written as
\begin{align}
   \nabla\cdot(\rho \mb v)  &= 0, \\    -\mu \mb v  &= \nabla p , \qquad \\
   \rho \mb v \cdot \nabla \theta   &= \nabla\cdot(\lambda \nabla\theta) + \omega - \frac{K}{\beta}\theta,\\ 
 \rho \mb v \cdot \nabla Y   &= \frac{1}{Le}\nabla\cdot(\lambda \nabla Y) - \omega,  \\
   \rho(1+q\theta)=1, &\qquad \mu = \lambda= (1+q\theta)^n ,
\end{align}
where
\begin{equation}
    \omega = \frac{\beta^2 (1+q)^{1-n}}{2LeS^2}  \rho Y \exp\left[\frac{\beta(\theta-1)}{ 1+q (\theta-1)/(1+q)}\right]. \label{omega}
\end{equation}
Here, $Le$ is the Lewis number of the reactant with respect to the hot inert gas. The planar flame solution admits a self-similar description of the form
\begin{align}
    &v_x = A(y) x, \quad v_y = v_y(y), \quad \theta=\theta(y), \label{darcyself1} \\ &Y=Y(y), \quad 
    p= -\frac{\mc A}{2}(x^2-y^2) + \Pi(y) +\text{cst.},     \label{darcyself2}
\end{align}
where $A(y)$ is the non-dimensional strain rate (scaled by $S_L/\delta_L$), satisfying $A(+\infty)=\mc A$ and $A(-\infty)=m\mc A$. The quantity $1/\mc A = (\delta_m/\delta_L)^2$, with $\delta_m = \sqrt{D_{T,1}/A_1}$ the mixing-layer thickness, corresponds to the Damköhler number.
 Integration of the $x$-momentum and continuity equations yields
\begin{equation}
    A(y) = \frac{\mc A}{\mu} , \quad  \rho v_y = - \mc A \eta \quad \text{where} \quad\eta= \int_0^y \frac{\rho}{\mu} dy. \label{Aeta}
\end{equation}
The vorticity field $\gb\Omega$ is given by the expression
\begin{equation}
    \gb\Omega = -x \frac{dA}{dy} \, \mb e_z = \frac{\mc Ax}{\mu^2}\frac{d\mu}{dy} \, \mb e_z ,
\end{equation}
which implies that vorticity is confined to the region where $\theta$ varies. In the presence of volumetric heat losses, vorticity can persist in the burnt gas region behind the flame  due to local cooling. This cooled gas is  re-heated as it approaches the hot stream. The spatial extent of the vorticity region thus depends on the strain rate, decreasing as $\mc A$ increases. In terms of the weighted coordinate $\eta$, the governing equations for $\theta(\eta)$ and $Y(\eta)$ reduce to
\begin{align}
    -\mc A\, \eta \frac{d\theta}{d\eta} &= \frac{d}{d\eta}\left(\rho\frac{d\theta}{d\eta}\right) + \frac{\mu \omega}{\rho} -\frac{K}{\beta}\frac{\mu\theta}{\rho }, \label{thetaY1}\\ -\mc A\, \eta \frac{dY}{d\eta} &= \frac{1}{Le}\frac{d}{d\eta}\left(\rho\frac{dY}{d\eta}\right) - \frac{\mu \omega}{\rho }, \label{thetaY2}
\end{align}
subject to the boundary conditions
\begin{align}
    \theta\to 0, \quad Y\to 1 \quad \text{as} \quad \eta\to +\infty, \label{BC1} \\
     \theta\to 1, \quad Y\to 0 \quad \text{as} \quad \eta\to -\infty. \label{BC2}
\end{align}

Once $\theta(\eta)$ and $Y(\eta)$ are determined by solving~\eqref{thetaY1}-\eqref{BC2}, various physical quantities of interest can be evaluated. These include the burning rate $\dot m$ per unit flame area (scaled by $S_L$), the flame temperature $\theta_f$, the fuel leakage $Y_f$,  the flame position $y=y_f$,  and the displacement thicknesses $\delta_\pm$. These quantities are defined by
\begin{align}  
    &\dot m= \int_{-\infty}^{+\infty}\omega dy, \qquad \theta_f=\theta(y_f), \label{four1} \\ &Y_f=Y(y_f), \qquad y_f = \frac{\displaystyle\int_{-\infty}^{+\infty} y \omega \, dy}{\displaystyle\int_{-\infty}^{+\infty} \omega \, dy} ,  \label{four2}
\end{align}
where the integrals over $y$ can be transformed to integrals over $\eta$ using $d\eta = dy \, \rho/\mu$.  The flame position, $y_f$, denoting the location of the reaction zone, is identified here as the mean location of the chemical activity. Due to both viscosity and density variations within the flame region, the normal flow in the far field gets displaced, which can be quantified using displacement thicknesses $\delta_\pm$, which are defined by
\begin{align} \label{dp}
    \delta_+ = \int_0^{+\infty} \left(1-\frac{\rho}{\mu}\right)dy  
\end{align}
so that $v_y \to -\mc A \, (y-\delta_+)$ as $y \to +\infty$ and
\begin{equation} \label{dm}
    \delta_- = \int_0^{-\infty} \left(1-\frac{r\rho}{m\mu}\right)dy 
\end{equation}
so that $v_y \to -m \mc A \, (y-\delta_-)$ as $y \to -\infty$

It is worth highlighting the influence of variable transport properties within this framework, specifically that of the local kinematic viscosity  $\mu/\rho$. This ratio is important in determining the stretched coordinate $\eta$ and defining the normal mass flux $\rho v_y=-\mc A \eta$. To put its effect in perspective, this ratio increases across the flame by a factor 
\begin{equation} \label{eq:resilience}
    \frac{\mu_2/\rho_2}{\mu_1/\rho_1} =\frac{r}{m}\approx  20 \qquad \text{when} \qquad q=5.
\end{equation}
This 20-fold increase drastically distorts the mapping between the physical coordinate $y$ and the weighted coordinate $\eta$. Thus, while in the unburnt gas region, $\rho v_y \approx -\mc A y + $ const., in the burnt gas region, $\rho v_y \approx -\mc A y/20 + $ const. This substantial reduction in slope on the hot side indicates a strong resistance to changes in the normal mass flux within the burnt gases. Alternatively, it is equivalent to a reduction in the tangential velocity component across the flame, leading to augmentation of streamline refraction, as discussed in~\cite{rajamanickam2026flame}.

The above formulation invites an illuminating comparison with the Navier--Stokes stagnation‑point flow, which is described in the Appendix. In the latter, an exact analytical expression for $A(y)$  is generally not available. However, recent work~\cite{weiss2018novel} showed that using a viscosity‑weighted coordinate $\eta'= \int_0^ydy/\mu$ provides a better approximation for the flow field (e.g., $\rho v_y\approx -\mc A\eta'$) than the standard density‑weighted Howarth--Dorodnitsyn coordinate $\eta''= \int_0^y\rho dy$, which is often adopted in the literature. In the present Darcy‑law configuration, the quantity that governs the flow is not $\mu$ alone but rather the ratio $\mu/\rho$, which plays the role that $\mu$ plays in a Navier--Stokes mixing layer. However, a fundamental distinction exists: in Navier--Stokes flows, using $\mu$ in the coordinate is a clever trick to improve numerical accuracy, whereas in Darcy flows, using $\mu/\rho$ in the coordinate is a mathematical necessity derived directly from the underlying physics.

A further point of observation concerns the behaviour of the strain rate across the flame. Specifically,
\begin{align}
    &\text{Darcy's law:} \quad \frac{A(-\infty)}{A(+\infty)} = m<1\qquad \text{and} \\  &\text{Navier--Stokes:} \quad \frac{A(-\infty)}{A(+\infty)} = \sqrt{r}>1 .
\end{align}
Thus, in the Darcy-law flame, the strain rate decreases from the unburnt to the burnt side, whereas in a classical Navier–Stokes stagnation-point flow, the opposite trend holds. In the Navier–Stokes case, the lighter burnt gas accelerates, steepening the velocity gradient, whereas in the Darcy case, the increased viscosity of the burnt gas leads to a gentler strain rate on the hot side. This contrast further underscores that the role of transport properties in each framework is fundamentally distinct.

\section{Numerical results}
Due to the temperature difference between the opposing cold and hot streams, a non-trivial  solution exists across all values of the strain rate $\mc A$ and heat-loss parameter $K$. However, these solutions do not always represent a usual, self-sustaining flame, as certain branches feature negligible burning rates and weak chemical activity. Furthermore, the governing equations admit multiple solutions for a fixed $\mc A$, leading to distinct ignition and extinction regimes which are highly sensitive to the mixture's transport properties.

We shall now present numerical results obtained by integrating equations~\eqref{thetaY1}-\eqref{BC2} with fixed parameters $\beta=10$ and $q=5$, while varying the strain rate $\mc A$, the heat-loss parameter $K$ and the Lewis number $Le$. The scaled laminar flame speed $S = S_L/S_L^0$ in~\eqref{omega} takes the values $S = \{0.9881, 0.9702, 0.9531, 0.9292, 0.8930, 0.8608, 0.8052\}$, corresponding to $Le = \{0.3, 0.5, 0.7, 1, 1.5, 2, 3\}$ when $\beta = 10$ and $q = 5$.

\begin{figure}[h!]
\centering
\includegraphics[scale=0.3]{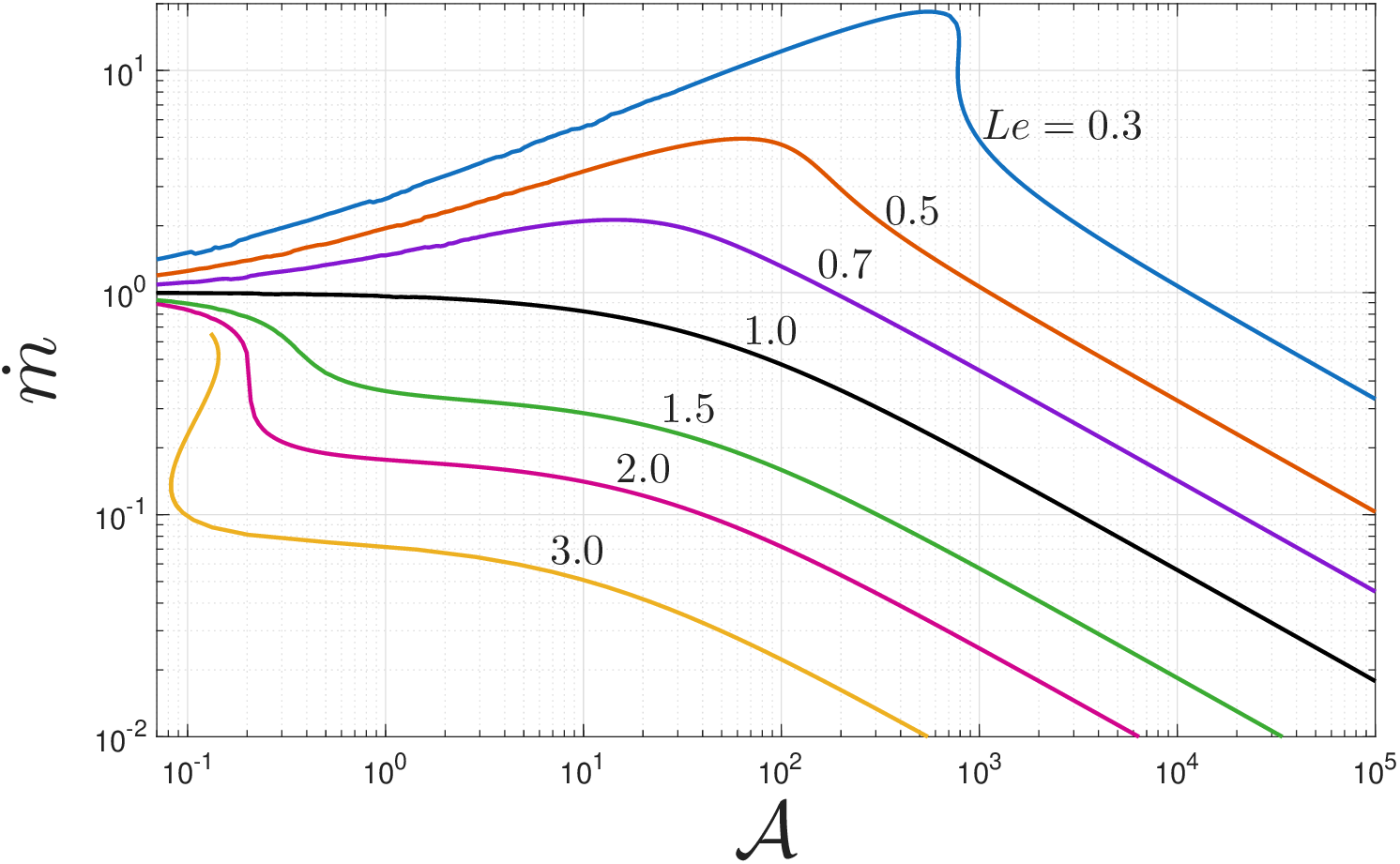} 
\includegraphics[scale=0.3]{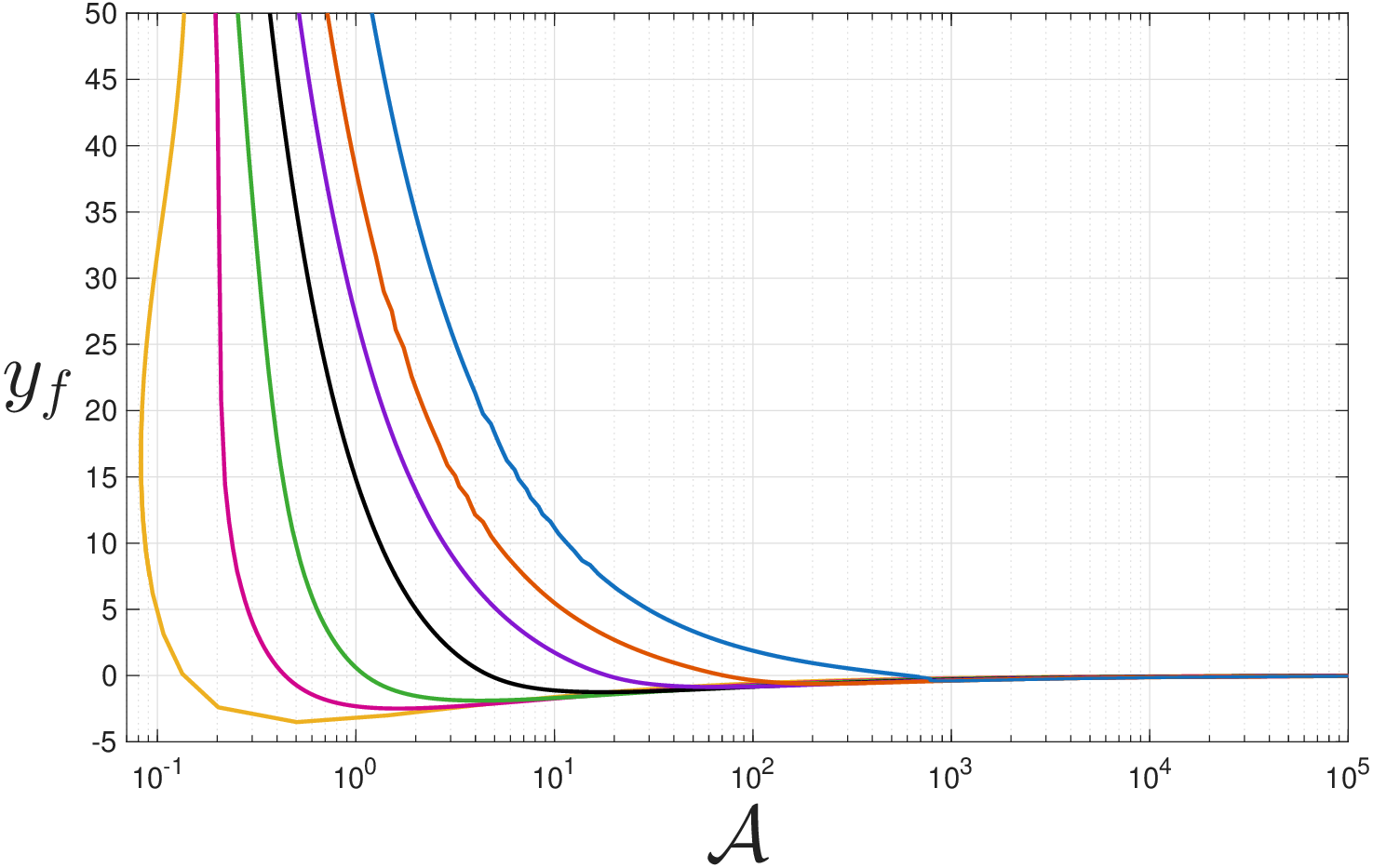} 
\caption{Burning rate  $\dot{m}$ (top) and flame position  $y_f$ (bottom) versus strain rate for various Lewis numbers under the adiabatic condition ($K=0$).}
\label{fig:adiabatic}
\end{figure}

\subsection{Adiabatic flames}

To isolate the purely hydrodynamic and transport-driven effects, we first examine the adiabatic regime $K=0$. Figure~\ref{fig:adiabatic} captures the dependence of the scaled burning rate $\dot m$ and flame position $y_f$ on $\mc A$ across a wide range of Lewis numbers ($0.3\leq Le \leq 3$).

At low strain rates $(\mc A\ll 1)$,  the flame resides well above the stagnation plane and far upstream in the unburnt mixture, where its structure closely resembles that of an unstrained, planar premixed flame, and $\dot m$ approaches its unstrained baseline value. As $\mc A$ increases, the flame migrates toward the stagnation plane, $y=0$,  and  crosses it at a moderate value of the strain rate. This ``moderate value" depends sensitively on $Le$.  For instance, crossing is delayed until intense strain rates of $\mc A \sim 10^2$ when $Le=0.3$, whereas it takes place much sooner, around $\mc A\sim 10^{-1}$, when $Le=3.0$.

At large values of the strain rate, the reaction zone migrates back toward the stagnation plane, approaching it from underneath (the hot-gas side). When the reaction zone resides beneath the stagnation plane, the local convective flow enters the flame from the burnt-gas side and exits toward the unburnt side; consequently, fuel transport to the reaction zone becomes heavily diffusion-dominated. The depth of the flame's excursion into the hot stream is strongly influenced by preferential diffusion and its impact on fuel delivery to the preheat zone. For sub-unity Lewis numbers $(Le<1)$, rapid mass diffusion enhances the fuel supply to the preheat zone, intensifying chemical consumption and preventing deep penetration into the hot-gas side. Conversely, for super-unity Lewis numbers ($Le>1$), sluggish fuel diffusion starves the reaction zone, pushing the flame deep into the hot stream where it must rely on elevated ambient temperatures to sustain itself.

Further examination of Figure~\ref{fig:adiabatic} reveals that the burning rate $\dot m$ approaches its unstrained baseline when $\mc A\ll 1$, but becomes vanishingly small at large strain rates. The functional dependence of $\dot m$ on $\mc A$ varies qualitatively with the Lewis number. For $Le<1$, the burning rate initially increases with the strain rate, reaching a maximum at moderate values of $\mc A$ before declining. A different trend is observed for $Le>1$, where $\dot m$ remains strictly below its unstrained value and decreases  with $\mc A$. This behaviour is directly tied to variations in the  flame temperature $\theta_f$; preferential diffusion renders the flame temperature super-adiabatic $(\theta_f>1)$ when $Le<1$ and sub-adiabatic $(\theta_f<1)$ when $Le>1$.

\begin{figure}[h!]
\centering
\includegraphics[scale=0.6]{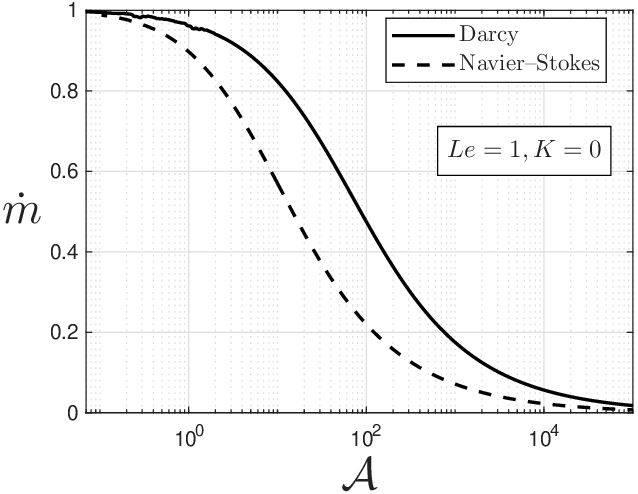} \hspace{1cm}
\includegraphics[scale=0.6]{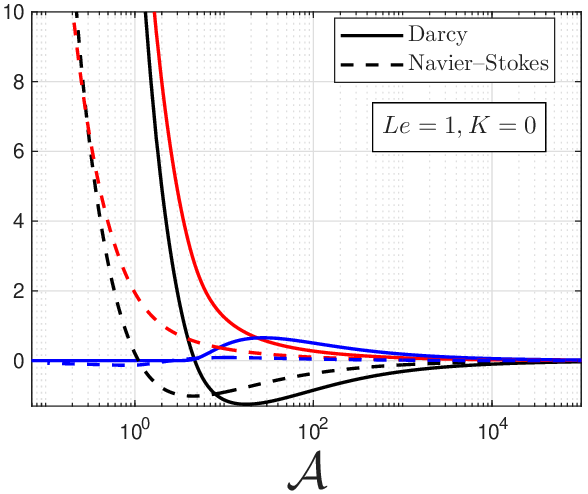} 
\caption{Comparison of Darcy's law (solid lines) and the Navier--Stokes equations (dashed lines) for $Le=1$ and $K=0$. The top panel plots the burning rate $\dot{m}$ as a function of $\mc A$, while the bottom panel shows the flame position $y_f$ (black) alongside the displacement thicknesses $\delta_+$ (red) and $\delta_-$ (blue).}
\label{fig:compare}
\end{figure}

Furthermore, the uniqueness of the solution depends on how much $Le$ deviates from unity. For near-unity and moderately non-unity Lewis numbers (such as $Le=0.5$ and $2.0$), $\dot m(\mc A)$ is uniquely defined across all strain rates. However, for more extreme deviations (such as $Le=0.3$ and $3.0$), the response curve exhibits a distinctive S-shaped ignition–extinction profile. Finally, we note that for low-Lewis-number fuels such as hydrogen, the flame is located far above the stagnation plane at ordinary values of $\mc A$; consequently, as seen in Figure~\ref{fig:adiabatic}, exceptionally high strain rates are required to keep it close to the stagnation plane and safely anchored within the mixing zone, thus preventing the flame from blowing out upwards.

\begin{figure}[h!]
\centering
\includegraphics[scale=0.6]{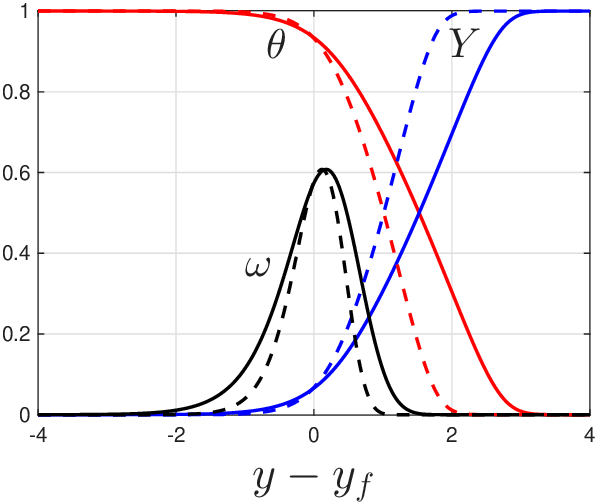}  
\includegraphics[scale=0.6]{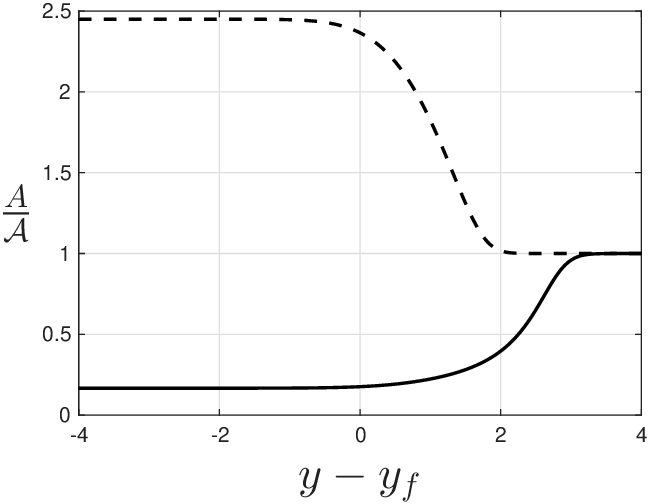}  
\includegraphics[scale=0.6]{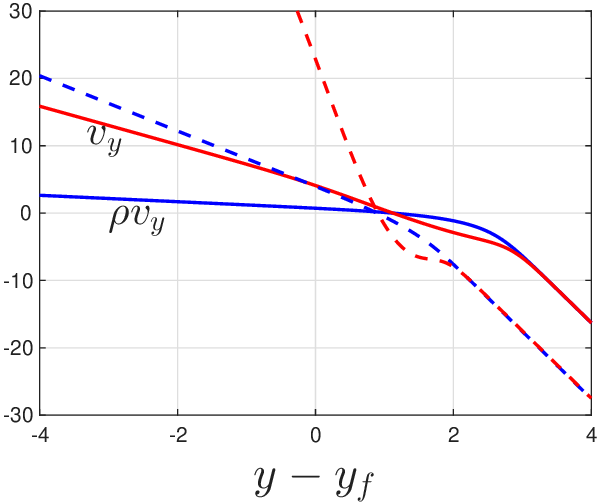} 
\caption{Comparison of flame structure and flow-field profiles between Darcy's law (solid lines) and the Navier--Stokes formulation (dashed lines) for $Le=1$, $\mc A=10$, and $K=0$. At this strain rate, the flame position settles at $y_f=-1.1219$ under Darcy's law and $y_f=-0.90475$ under the Navier--Stokes equations.}\label{fig:compare2}
\end{figure}

\begin{figure*}[h!]
\centering
\includegraphics[scale=0.5]{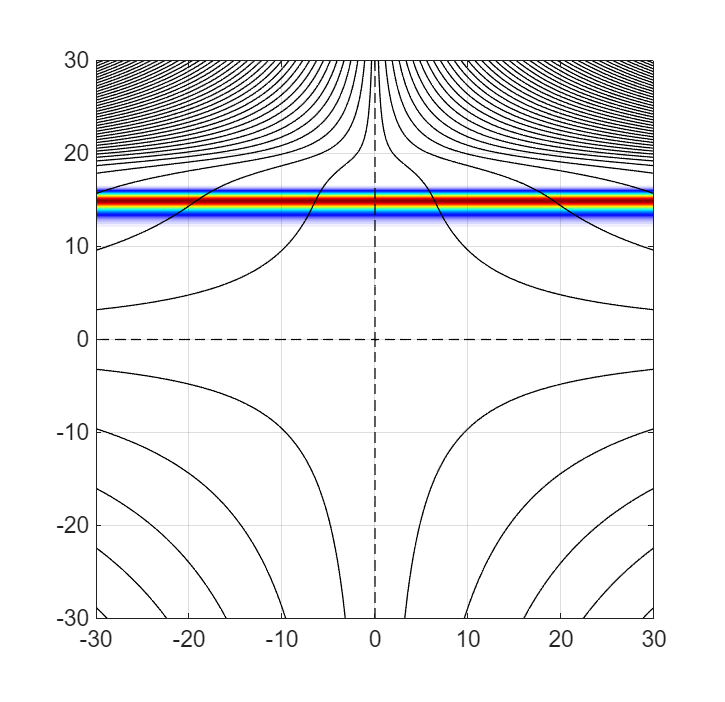} 
\includegraphics[scale=0.5]{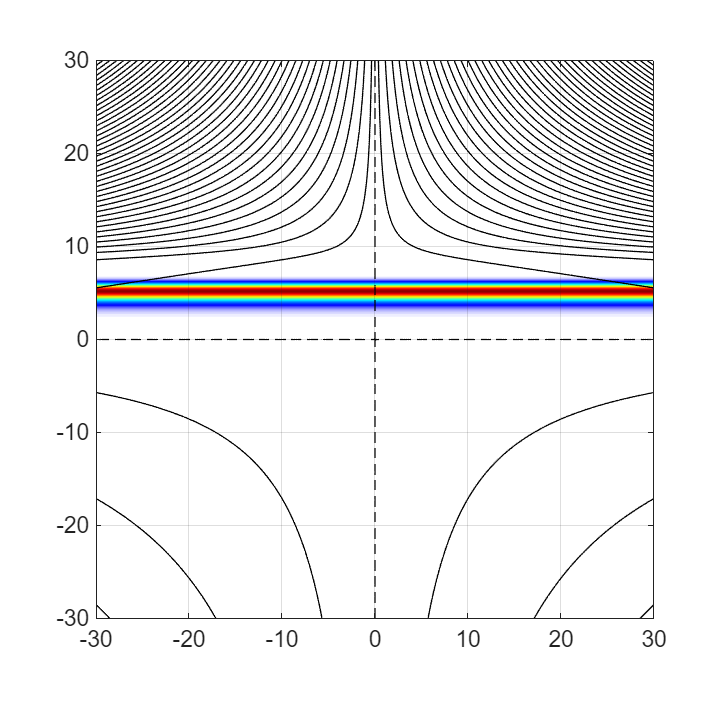} 
\includegraphics[scale=0.5]{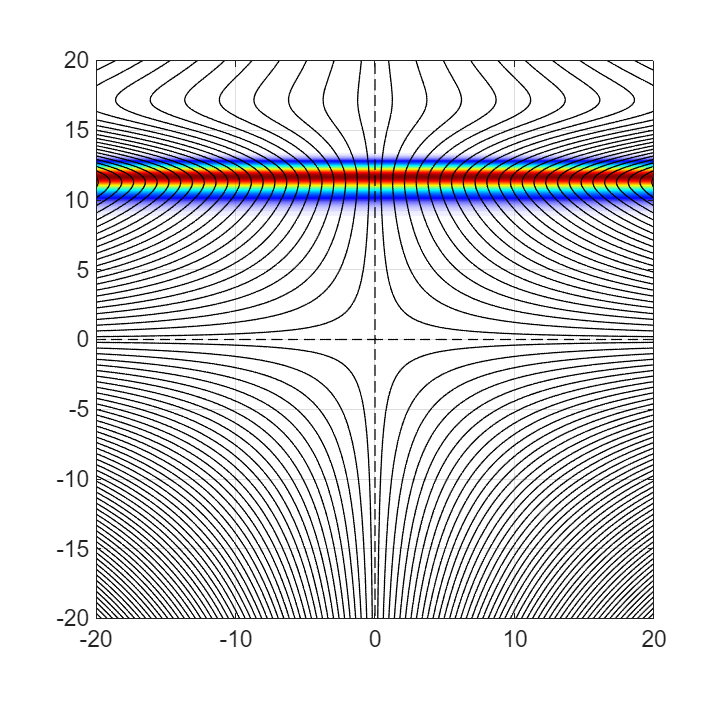} 
\includegraphics[scale=0.5]{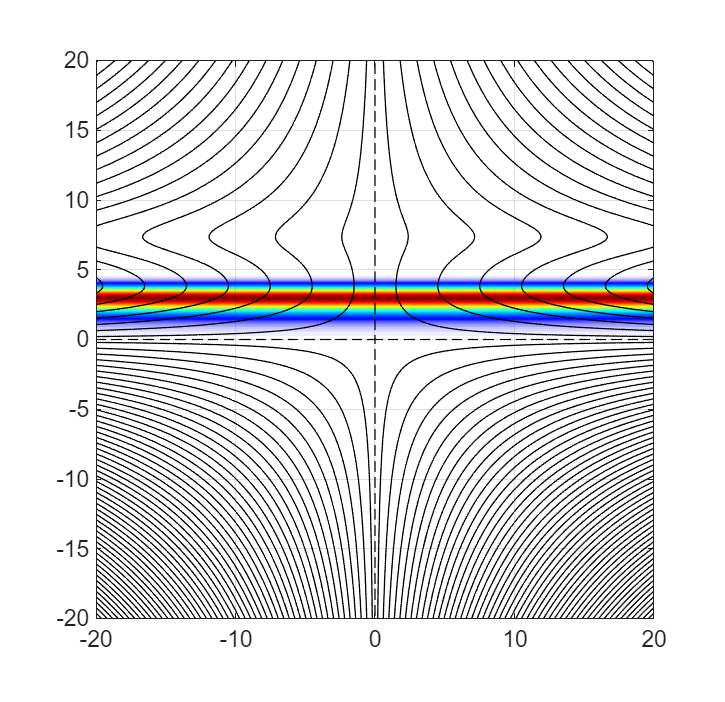} 
\caption{Reaction-rate fields (in colour) and streamlines for $Le=1$ and $K=0$. The top panels correspond to  Darcy's law at $\mc A=1$ (left) and $\mc A=2$ (right); the bottom panels correspond to  the Navier--Stokes equations at $\mc A=0.2$ (left) and $\mc A=0.5$ (right).} \label{fig:strreamline}
\end{figure*}

\subsection{Comparison between Navier--Stokes and Darcy's stagnation point flows}

In this subsection, we compare the numerical results obtained under Darcy's law with those computed from the classical Navier--Stokes formulation, which is outlined in the Appendix for the convenience of the reader. Figures~\ref{fig:compare} and~\ref{fig:compare2} illustrate this comparison for $Le=1$ and $K=0$, where solid lines denote the Darcy-law solution and dashed lines represent the Navier--Stokes solution.

The top panel of Figure~\ref{fig:compare} displays the scaled burning rate $\dot{m}$ as a function of the strain rate $\mc{A}$. Across all strain rates, $\dot{m}$ is consistently higher under Darcy's law than under the Navier–Stokes formulation. This enhancement is intrinsically linked to the distinct flow-field modifications that occur across the flame zone under each hydrodynamic framework. In particular, the twentyfold increase in kinematic viscous resistance, highlighted in the text following~\eqref{eq:resilience}, acts to dampen the local flow field, and hence tends to resist flame migration toward the stagnation plane. As a result, under a given intense strain rate, the Darcy flame offers greater resistance to being pushed toward the stagnation plane, maintaining a larger standoff distance than its more compressed Navier–Stokes counterpart, as seen in the bottom panel.

The three panels of Figure~\ref{fig:compare2} describe the flame structure and the flow field for a fixed strain rate of $\mc A=10$. As shown in the bottom panel, the friction-dominated Darcy flow exhibits a significantly gentler slope in the normal mass flux $\rho v_y$ on the hot side compared to the inertia-driven Navier--Stokes counterflow. This difference demonstrates the deceleration of the flow on the burnt-gas side, which is characterised by higher viscosity. Furthermore, the friction-dominated framework significantly alters the flame's location and its thickness compared to the inertia-driven case.  In particular, the top panel illustrates the expanded flame thickness under Darcy's law. This behaviour stems from the reduced local strain rate, and the consequent increase in reactant residence time, on the burnt-gas side, as captured in the middle panel.

As a direct consequence of the strain rate variations across the flame, the flow patterns are fundamentally altered when comparing the Darcy framework to the Navier–Stokes description, as shown in Fig.~\ref{fig:strreamline}. These distinct flow structures provide key insights into designing experimental setups to sustain a premixed flame at a target location. For instance, under Darcy's law, the oncoming stream of unburnt gas must be sufficiently accelerated relative to the burnt gas stream. This operational requirement stands in striking contrast to the classical Navier--Stokes configuration, where it is instead the burnt-gas stream that requires a higher acceleration.

\subsection{Effects of conductive heat losses}

Finally, we examine the influence of volumetric heat losses ($K \neq 0$) on the flame structure and extinction characteristics. The analysis is presented for three values of the Lewis numbers, for simplicity. 

\begin{figure}[h!]
\centering
\hspace{-0.45cm}\includegraphics[scale=0.35]{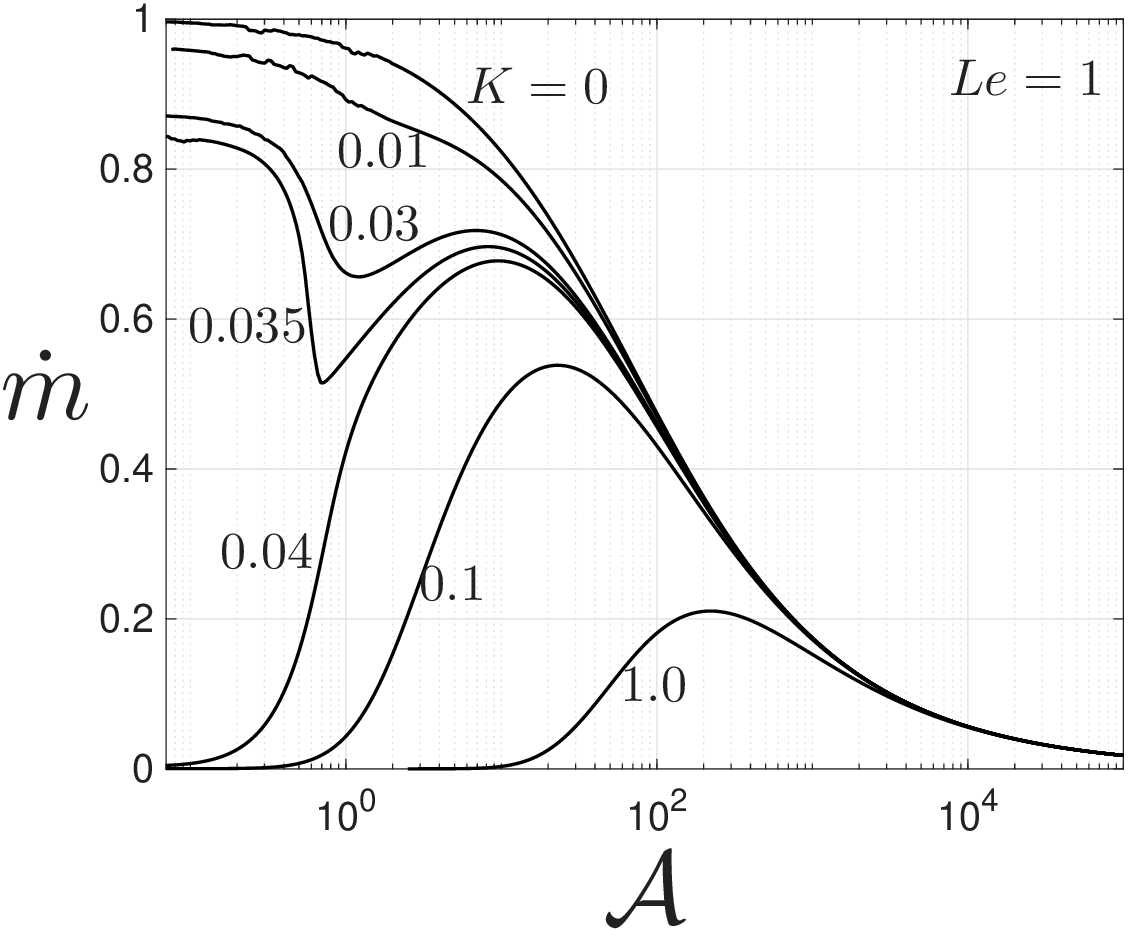} 
\caption{Scaled burning rate $\dot{m}$ as a function of the strain rate $\mc A$ for selected values of the heat-loss parameter $K$ at $Le=1$.}\label{fig:heatloss}
\end{figure}

In Fig.~\ref{fig:heatloss}, we plot the burning rate for selected values of $K$. At low strain rates, where the flame structure closely resembles that of an unstrained planar flame, the system exhibits a clear quenching threshold, specifically at $K > 0.035$ However, when the strain rate $\mc{A}$ is sufficiently large, a non-trivial solution persists regardless of the heat-loss intensity. This figure highlights the transition from the unstrained quenching limit to the strained flame regime;   the quenching boundary disappears for roughly $\mc{A} > 1$. Consequently, a strained flame can withstand severe thermal losses that would inevitably extinguish its unstrained counterpart.  

\begin{figure}[h!]
\centering
\includegraphics[scale=0.5]{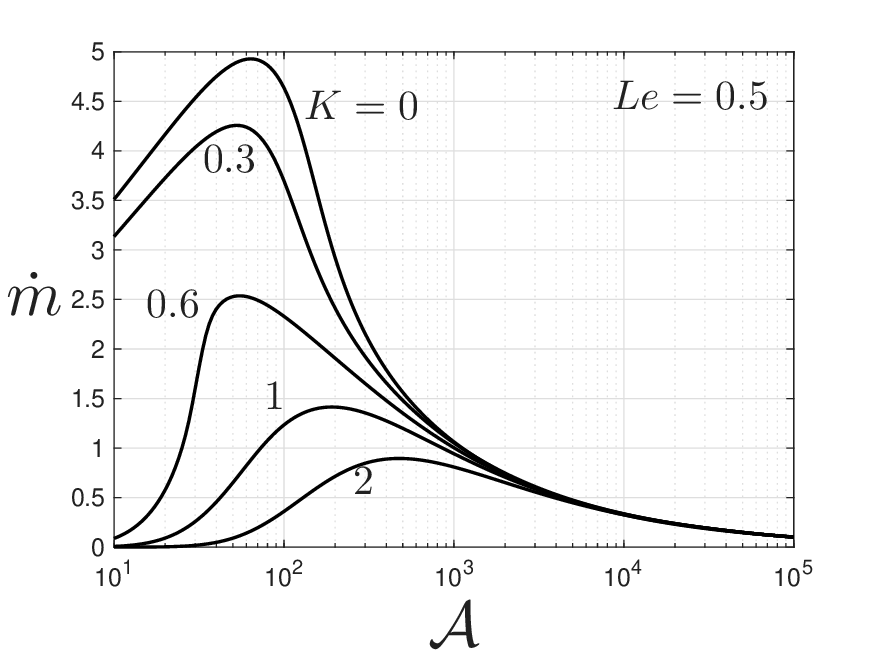} 
\includegraphics[scale=0.5]{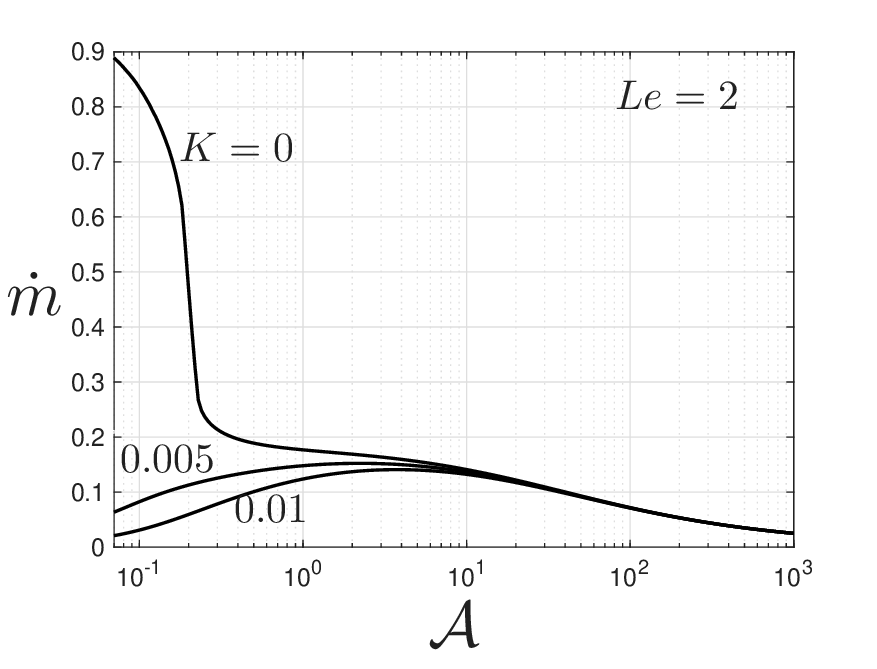} 
\caption{Scaled burning rate $\dot{m}$ as a function of the strain rate $\mc A$ for selected values of the heat-loss parameter $K$ at $Le=0.5$ (top) and $Le=2$ (bottom).}\label{fig:heatlossLe}
\end{figure}

Figure~\ref{fig:heatlossLe} illustrates the effects of heat losses for $Le=0.5$ and $Le=2$. The general trend is qualitatively similar to the $Le=1$ case; in particular, the non-monotonic behaviour of $\dot m$ versus $\mc A$ noted earlier in the adiabatic case persists when $K \neq 0$ in sub-unity Lewis number mixtures, and indeed for arbitrary Lewis numbers when $K$ exceeds a certain critical value. Furthermore, for any fixed value of the Lewis number, the $\dot m$ versus $\mc A$ curves collapse into a single curve as $\mc A \to \infty$, irrespective of the value of $K$. This collapse is expected because, in this limit, the flame is pushed towards the stagnation plane near the hot inert burnt gas, rendering it less sensitive to $K$. Moreover, while combustion is sustained with a significant burning rate in the presence of large heat losses for the sub-unity Lewis-number case, the $Le=2$ case remains highly susceptible to thermal losses, with its burning rate becoming negligibly small even at relatively low values of $K$. This highlights the coupled influence of preferential diffusion and heat loss. Consequently, this combined role suggests, for example, that one can reasonably expect hydrogen flames ($Le<1$) to be sustained under conditions where hydrocarbon flames ($Le>1$) cannot.

\section{Conclusions}

In this study, we have investigated the structure and stabilisation of nonadiabatic strained premixed flames governed by Darcy's law, relevant to strongly confined environments such as narrow Hele-Shaw burners and porous media. The work was carried out within the canonical configuration of a planar stagnation-point flow, and has unravelled a fundamental departure from classical results. Specifically, the spatial variation and jump in the strain rate across the flame zone are driven by a sharp variation in fluid viscosity rather than by the density variations characteristic of conventional inertia-dominated Navier--Stokes counterflows. The kinematic viscous resistance ($\mu/\rho\kappa$) varies drastically across the flame, effectively forcing the burnt-gas side to act as a strong viscous barrier that alters streamline refraction and modifies the normal mass flux.

Furthermore, analysing the coupled effects of non-unity Lewis numbers and volumetric heat losses has revealed unique burning characteristics. In particular, extreme deviations of \(Le\) from unity introduce an S-shaped ignition–extinction response, while the local strain variations enable the flame to withstand severe thermal losses that would otherwise extinguish its unstrained counterpart.

These localised stagnation-point findings carry direct implications for more complex, multi-dimensional flame topologies under Darcy's law. For example, a curved flame front experiences a localized aerodynamic strain rate roughly proportional to $U/R$, where $R$ represents the local radius of flame curvature and $U$ is a characteristic velocity scale. This velocity scale is governed either by the imposed external flow velocity or, in a quiescent medium, by the flame's own propagation velocity scaled by gas expansion. Thus, the extinction limits and burning rate modifications mapping out the $\dot{m}$--$\mathcal{A}$ curves here serve as a necessary baseline for predicting the localized burning rate over wrinkled front structures. This coupling between local strain and propagation speed aligns directly with our recent hydrodynamic stability analysis~\cite{daou2025hydrodynamic}, which shows that such interfaces are prone to Saffman--Taylor (viscous fingering) instabilities intimately tied to the propagation velocity. Future work will aim to extend this planar flame configuration to a Tsuji-type Hele-Shaw burner in the presence of an imposed flow. A key conclusion drawn from the present study is that the hot-gas injection through the Tsuji burner (a circular cylinder in a Hele-Shaw cell) does not need to be as strong as the incoming unburnt gas stream, owing to the significant viscous barrier effect identified herein. This configuration can be readily realised in narrow channels, making it a promising avenue for future experimental validation even in the presence of severe heat losses.

\section*{Acknowledgments} 

This work was supported by UK EPSRC through Grant No.~APP39756.

\bibliographystyle{elsarticle-num}
\bibliography{elsarticle-template}

\section*{Appendix: Formulation based on Navier--Stokes equations}

To evaluate the precise impact of Darcy-friction modifications on the counterflow field, this appendix presents the classical inertia-dominated Navier--Stokes formulation used as the reference benchmark throughout this study. The governing equations and corresponding boundary conditions for the self-similar solution are detailed below. 

The non-dimensionalisation of various variables are same as those given in~\eqref{eqn1}-\eqref{eqn3}, except that now the pressure field is scaled as $p_*=p/\rho_1 S_L^2$. After dropping the asterisk, the pressure field in the self-similar solution is given by
\begin{equation} \nonumber
    p = - \frac{\mc A}{2}(x^2+y^2) + \Pi(y) + \text{const.} , \label{NSself}
\end{equation}
which replaces the expression in~\eqref{darcyself2}. The governing equations are given by~\cite{linan1993autoignition}
\begin{align*}
    \rho A + \frac{d(\rho v_y)}{dy} &=0,\\
    \rho A^2 + \rho v_y \frac{dA}{dy} &= \mc A^2 + Pr \frac{d}{dy}\left(\mu \frac{dA}{dy}\right),\\
    \rho v_y \frac{d\theta}{dy} &= \frac{d}{dy}\left(\lambda\frac{d\theta}{dy}\right) + \omega - \frac{K}{\beta}\theta,\\
    \rho v_y \frac{dY}{dy} &= \frac{1}{Le}\frac{d}{dy}\left(\lambda\frac{dY}{dy}\right) - \omega,\\
    \rho(1+q\theta)=1, &\qquad \mu = \lambda= (1+q\theta)^n,
\end{align*}
where $Pr=\mu_1/\rho_1D_{T,1}\approx 0.7$ is the Prandtl number. The boundary conditions are given by $v_y(0)=0$,
\begin{align*}
    &A \to \mc A, \quad \theta \to 0, \quad Y \to 1 \quad \text{as} \quad y \to +\infty,\\
    &A \to \sqrt{r}\mc A, \quad \theta \to 1, \quad Y \to 0 \quad \text{as} \quad y \to -\infty.
\end{align*}
The pressure field is determined by integrating the $y$-momentum equation,
\begin{align*}
    \frac{d\Pi}{dy}= -\rho v_y \frac{dv_y}{dy} + \mc A^2 y + Pr \left\{\mu \frac{dA}{dy} + \frac{d}{dy}\left[\mu \left(\frac{4}{3}\frac{dv_y}{dy} - \frac{2}{3}A\right)\right]\right\},
\end{align*}
subject to the condition $\Pi(+\infty)=0$. While the four physical quantities defined in~\eqref{four1}-\eqref{four2}  still retain the same definition, the displacement thicknesses $\delta_\pm$ given in~\eqref{dp}-\eqref{dm} need to be redefined as
\begin{equation} \nonumber
    \delta_+ = \int_0^{+\infty} \left(1-\frac{\rho A}{\mathcal A}\right)dy
\end{equation}
so that $v_y \to -\mc A \, (y-\delta_+)$ as $y \to +\infty$ and
\begin{equation} \nonumber
    \delta_- = \int_0^{-\infty} \left(1-\frac{\sqrt{r}\rho A}{\mathcal A}\right)dy
\end{equation}
so that $v_y \to -\sqrt{r}\mc A \, (y-\delta_-)$ as $y \to -\infty$.

\end{document}